\documentclass[11pt,namedreferences]{kluwer}
\usepackage[dvips]{graphicx}

\def\edcomment#1{\iffalse\marginpar{\raggedright\sl#1\/}\else\relax\fi}
\reversemarginpar

\begin{document}
\begin{article}
\begin{opening}
\title{Formation of the Planetary Sequence in a Highly \\  
       Flattened Disk of Frequently Colliding Planetesimals} 
\author{Evgeny Griv, Michael Gedalin, Edward Liverts, David Eichler}
\institute{Dept. of Physics, Ben-Gurion University, Beer-Sheva 84105,
       Israel}
\author{Chi Yuan}
\institute{Academia Sinica Institute of Astronomy,
       Taipei 11529, Taiwan}

\begin{abstract}
The kinetic theory is used to study the evolution of the
self-gravitating disk of planetesimals.  The effects of frequent
collisions between planetesimals are taken into account by using
a Krook integral in the Boltzmann kinetic equation.  It is shown 
that as a result of an aperiodic collision-dissipative
instability of small gravity disturbances the disk is subdivided
into numerous dense fragments.  These can eventually condense
into the planetary sequence.

\end{abstract}
\end{opening}
Solar system formation is thought to start with dust particles
settling to the central plane of a nebula to form a thin dust
layer around the equatorial plane.  During the early evolution 
of such a rapidly rotating disk it is believed that the dust
particles coagulate into numerous kilometer-sized rocky bodies 
(planetesimals).  See Taylor (1992) as a review of the problem. 

In a swarm of planetesimals direct physical collisions inevitably
become an important factor (Taylor 1992).  One can suggest that
planets (Mercury, Venus, $\dots$, Neptune) accreted subsequently
from a hierarchy of colliding planetesimals.  (The combination of
low mass with a highly inclined and eccentric orbit is a major
reason for not according Pluto planetary status: observations point
to the orbit of Neptune as the true outer boundary of the planetary
system.)

Our principal idea is to regard the formation of the planetary system
as a possible last stage in the formation of the self-gravitating,
highly flattened solar nebula with frequent, almost elastic collisions
between rocky planetesimals.  We argue that a collision-dissipative
collective instability of small-amplitude gravity perturbations
developing in such a disk leads to the formation of an arrangement
of dense aggregates.  We speculate that subsequent substantial
gravitational interaction between these dense aggregates circling
the primordial sun would result in the formation of the planetary 
sequence.  When cores of the giant planets reach a critical mass
($\sim 10$ masses of the Earth) they begin to accrete the 
interplanetary gas.

The collision motion of an ensemble of identical planetesimals in 
the plane, in the frame of reference rotating with angular velocity 
$\Omega$, can be described by the Boltzmann kinetic equation
\begin{displaymath}
\frac{\partial f}{\partial t} + \vec{v} \cdot \frac{\partial f}
{\partial \vec{r}} - \frac{\partial \Phi}{\partial \vec{r}} \cdot
\frac{\partial f}{\partial \vec{v}} = \left( \frac{\partial f}
{\partial t} \right)_{\mathrm{coll}} , 
\end{displaymath}
where $f ({\vec r},{\vec v},t)$ is the phase-space distribution 
function of planetesimals, $\Phi (\vec{r},t)$ is the total
gravitational potential, and $(\partial f/\partial 
t)_{\mathrm{coll}}$ is the so-called collision integral which 
takes into account effects due to the discrete-point nature of 
the gravitational charges and defines the change of $f$
arising from ordinary interparticle collisions. 
There are only elastic physical collisions, and momentum is
conserved in collisions.  There is no correlation in motion between
the colliding species, that is, Boltzmann's hypothesis of molecular
chaos is adopted.

The Poisson equation in a suitable form is $\nabla^2 \Phi = 4\pi 
G\sigma \delta (z)$, where $\sigma ({\vec r},t)$ is the surface
mass density.  The Boltzmann and Poisson equations,
with the distribution function of planetesimals and
appropriate boundary conditions, give a complete description of the
problem for infinitesimally thin disk modes in a self-consistent 
field problem.  In the present study, this system of equations is
simplified by considering the lowest WKB approximation (in the 
limit $\nu_{\mathrm{c}} \gg \Omega$, where $\nu_{\mathrm{c}}$ is
the frequency of collisions between planetesimals); this is
accurate for short wave perturbations only, but qualitatively
correct even for perturbations with a longer wavelength, of the
order of the system radius.

As is known, the Boltzmann equation is nearly intractable because
of the complicated collision integral.  This integral can be
approximated in various ways.  In this work we use the simplified
kinetic model when the exact integral $(\partial f/\partial t)_
{\mathrm{coll}}$ is replaced by a phenomenological term in the 
form of a Krook model (Lifshitz \& Pitaevskii 1981; Griv \& Peter 
1996; see also Griv et al. 1999).

Perturbations in the gravitational field cause perturbations 
to the planetesimal distribution function.  In the linear 
approximation, one can therefore write $f(\vec{r},\vec{v},t)=f_0
(r,\vec{v})+f_1(\vec{r},\vec{v},t)$, 
where $|f_1| \ll f_0$ and $f_1$ is a
function rapidly oscillating in space and time.  We consider zeroth
order disk inhomogeneity in the $r$-direction only.  Initially the
disk is in equilibrium, $\partial f_0 /\partial t=0$.  The function
$f_0$ describes the rotating ``background" against which small
perturbations develop.  If an initial perturbation grows,
the system is called unstable.

As a result of this study, the dispersion relation connecting the
Doppler-shifted wavefrequency $\omega_*$ and the wavenumber $k$ of
perturbations is found:
\begin{displaymath}
1 - \frac{2 \pi G \sigma_0}{|k| c^2}
\left\{ 1 - \frac{\sqrt{\pi} g}{\sqrt{2} k c} \exp
\left( \frac{g^2}{2 k^2 c^2} \right) \left[ 1 - \mbox{erf}
\left( {g \over \sqrt{2} k c} \right) \right] \right\} =0 ,
\end{displaymath}
where $\sigma_0 (r)$ is the equilibrium surface density, 
$c$ is the dispersion of random velocities of planetesimals 
(``temperature"), $g=i(\gamma + \nu_{\mathrm{c}})$, and 
$\mbox{erf} (\beta)$ denotes the error function.  The
wavefrequency is represented in the form  $\omega \equiv 
i \gamma$, where $\gamma$ is real and positive (cf. Binney
\& Tremaine 1987, p. 292).

A numerical solution of the dispersion relation is shown in 
Figure 1.  As follows from this solution, the dispersion 
relation gives the condition for an aperiodic
collision-dissipative instability ($\gamma > 0$):
\begin{displaymath}
\lambda > \lambda_{\mathrm{crit}} \approx c^2/G \sigma_0 
\quad \mathrm{and} \quad \lambda = 2 \pi / k ,
\end{displaymath}
which is just the familiar Jeans criterion for a nonrotating 
medium and $\lambda_{\mathrm{crit}}$ is the ordinary Jeans
length (Binney \& Tremaine 1987).  Frequent collisions thus 
remove the rotational stabilization in a flat system.  Thus, 
the theory predicts that a gravitating disk with frequent
interparticle collisions is aperiodically unstable for
perturbations with wavelengths greater than the Jeans length.

At the wavelength $k_* \approx G \sigma_0 / c^2$ the growth 
rate of the instability is maximum (Fig. 1).  It means that of
all harmonics of initial perturbation, one perturbation with the
maximum of the growth rate and with $k_*$ will be formed
asymptotically in time.  The numerous condensations with 
typical dimensions and distances between them $\lambda_* = 2\pi
/ k_*$ that arise will remain localized in space and grow,
because the instability will be an aperiodic.

\begin{figure}
\centerline{\includegraphics[width=\textwidth,keepaspectratio]
{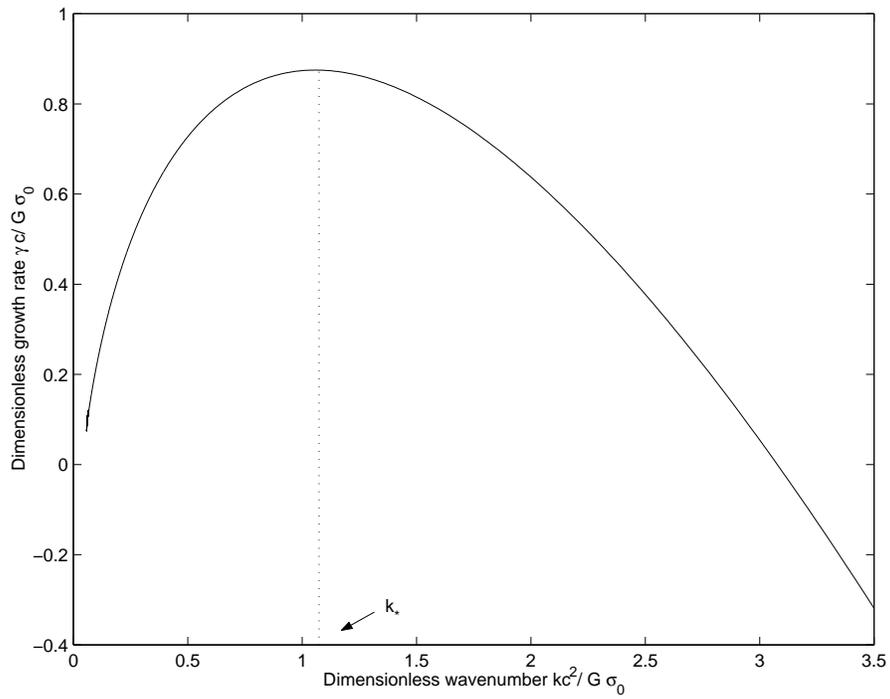}}
\caption{The dispersion relation for reasonable parameters of a
two-dimensional system of frequently colliding planetesimals.}
\end{figure}

\subparagraph{Acknowledgments}

Support from the Israel Science Foundation and the Israeli 
Ministry of Immigrant Absorption is acknowledged.

\end{article}
\end{document}